# Fabrication of low-loss quasi-single-mode PPLN waveguide and its application to a modularized broadband high-level squeezer

Takahiro Kashiwazaki,[1),a)] Taichi Yamashima,[2)] Naoto Takanashi,[2)] Asuka Inoue,[1)] Takeshi Umeki,[1)] and Akira Furusawa[2),3),b)]

[1)] *NTT Device Technology Labs, NTT Corporation, 3-1, Morinosato Wakamiya, Atsugi, Kanagawa, 243-0198, Japan*

[2)] *Department of Applied Physics, School of Engineering, The University of Tokyo, 7-3-1 Hongo, Bunkyo-ku, Tokyo, 113-8656, Japan*

[3)] *Optical Quantum Computing Research Team, RIKEN Center for Quantum Computing, 2-1 Hirosawa, Wako, Saitama, 351-0198, Japan*

[a)] Author to whom correspondence should be addressed: takahiro.kashiwazaki.dy@hco.ntt.co.jp

[b)] Electronic mail: akiraf@ap.t.u-tokyo.ac.jp

# Fabrication of low-loss quasi-single-mode PPLN waveguide and its application to a modularized broadband high-level squeezer

Takahiro Kashiwazaki,[1),a)] Taichi Yamashima,[2)] Naoto Takanashi,[2)] Asuka Inoue,[1)] Takeshi Umeki,[1)] and Akira Furusawa[2),3),b)]

[1)] *NTT Device Technology Labs, NTT Corporation, 3-1, Morinosato Wakamiya, Atsugi, Kanagawa, 243-0198, Japan*

[2)] *Department of Applied Physics, School of Engineering, The University of Tokyo, 7-3-1 Hongo, Bunkyo-ku, Tokyo, 113-8656, Japan*

[3)] *Optical Quantum Computing Research Team, RIKEN Center for Quantum Computing, 2-1 Hirosawa, Wako, Saitama, 351-0198, Japan*

[a)] Author to whom correspondence should be addressed: takahiro.kashiwazaki.dy@hco.ntt.co.jp

[b)] Electronic mail: akiraf@ap.t.u-tokyo.ac.jp

**ABSTRACT**

A continuous-wave (CW) broadband high-level optical quadrature squeezer is essential for high-speed large-scale fault-tolerant quantum computing on a time-domain-multiplexed continuous-variable optical cluster state. CW THz-bandwidth squeezed light can be obtained with a waveguide optical parametric amplifier (OPA); however, the squeezing level have been insufficient for applications of fault-tolerant quantum computation because of degradation of the squeezing level due to their optical losses caused by the structural perturbation and pump-induced phenomena. Here, by using mechanical polishing processes, we fabricated a low-loss quasi-single-mode periodically-poled $LiNbO_3$ (PPLN) waveguide, which shows 7% optical propagation loss with a waveguide length of 45 mm. Using the waveguide, we assembled a low-loss fiber-pigtailed OPA module with a total insertion loss of 21%. Thanks to its directly bonded core on a $LiTaO_3$ substrate, the waveguide does not show pump-induced optical loss even under a condition of hundreds of milliwatts pumping. Furthermore, the quasi-single-mode structure prohibits excitation of higher-order spatial modes, and enables us to obtain larger squeezing level. Even with including optical coupling loss of the modularization, we observe 6.3-dB squeezed light from the DC component up to a 6.0-THz sideband in a fully fiber-closed optical system. By excluding the losses due to imperfections of the modularization and detection, the squeezing level at the output of the PPLN waveguide is estimated to be over 10 dB. Our waveguide squeezer is a promising quantum light source for high-speed large-scale fault-tolerant quantum computing.

Measurement-based quantum computing (MBQC) on a time-domain-multiplexed continuous-variable optical cluster state has great potential for high-speed large-scale fault-tolerant quantum computing.[1-3] To ensure this potential, a continuous-wave (CW) broadband high-level optical quadrature squeezer is desired as a quantum light source to generate quantum entangled states[1,4,5] and ancillary input quantum states.[6] Broadband CW squeezed light allows us to achieve high-clock-rate signal processing.[1] It also allows us to generate a large-scale time-domain-multiplexed optical cluster state with the use of short-length optical delay-line interferometers.[4,5] Furthermore, high-level squeezed light realizes high-fidelity continuous-variable quantum gates[7] and high-dimensional quantum entanglement.[2,8]

One of the most promising way to obtain CW broadband squeezed light is to use a single-pass optical parametric amplifier (OPA) instead of a cavity-based optical parametric oscillator (OPO). A single-pass OPA inherently shows a terahertz (THz)-order bandwidth because of its no-cavity configuration.[9-12] Recently, various waveguide OPAs show high gain parametric process thanks to its large light-confining-effect along the waveguide.[13,14] In particular, a second-order nonlinear ($\chi^{(2)}$) waveguide OPA shows a highly efficient broadband optical parametric process even for CW light.[15-19] In 2011, a periodically poled LiNbO$_3$ (PPLN) waveguide achieved 4600-%/W internal second-harmonic-generation.[19] In addition, over-30-dB phase-sensitive amplification for a CW signal was successfully demonstrated in 2019.[16] The PPLN waveguide OPA also recorded CW 6.3-dB squeezing with 2-THz spectral bandwidth in 2020,[17] whose single-mode condition prohibits contamination of anti-squeezed noise from higher order modes. However, the reported squeezing levels with the $\chi^{(2)}$ waveguide OPAs are still insufficient for various quantum operations, such as quantum error correction.[20-23] The reason for the low squeezing level is that it has been difficult to achieve low optical loss and high $\chi^{(2)}$ nonlinearity in the same time. Therefore, for a $\chi^{(2)}$ waveguide squeezer, increasing the squeezing level is still a challenge. Moreover, a fiber-pigtailed squeezer is preferable because of its compatibility with various reliable optical components,[24,25] which are required for MBQC using time-domain multiplexing technique with optical-fiber-based delay-line interferometers.[4,5]

Requirements for a $\chi^{(2)}$ waveguide core achieving a higher squeezing level are low optical loss for squeezed light, high-pump-power durability to keep the low loss, and higher-order-mode suppression. Optical loss causes an invasion of vacuum noise, which deteriorates the squeezing level.[12] Durability against strong pumping is required because, in a nonlinear crystal, high-power-pump light causes various phenomena, such as a photorefractive effect causing light scattering[26,27] and photo-induced infrared light absorption.[28,29] In addition, it is important to suppress excitation of higher-order modes because their anti-squeezed noise reduces the observed squeezing level. In Ref. 17, with a use of a single-mode PPLN waveguide fabricated by dry etching, 6.3 dB of squeezing was measured without any pump-induced deterioration even with over-100-mW CW pump light. Its bonded and ZnO-doped waveguide core allowed higher power pump injection[17,18,30,31] compared to another waveguide fabricated by ion-exchange or the metal diffusion method.[32] However, the optical loss of a dry-etched waveguide tends to be high due to its rough sidewalls. Generally, dry-etched waveguides show about 0.3 dB/cm of optical transmission loss,[33,34] which corresponds to 25% of optical loss for a 45-mm-long waveguide. This is because, during the dry-etching process, byproducts redeposit on the sidewalls and increase the surface roughness.[34-37] This has been a long-standing problem for the dry etching of LN waveguides.

The use of a mechanically sculptured waveguide with a directly bonded core is one solution to increase the squeezing level, because the waveguide has smooth sidewalls polished with a diamond abrasive during the fabrication process.[18,38-41] In this work, by using the mechanically sculpturing process, we fabricated a quasi-single-mode ZnO-doped PPLN waveguide, whose waveguide core is directly bonded onto a LiTaO$_3$ substrate. With a length of 45 mm, the PPLN waveguide shows optical loss of 7%. On top of that, since the quasi-single-mode waveguide allows only the existence of two modes for squeezed light, the anti-squeezed components of higher order modes become negligibly small. Using the fabricated waveguide assembled as a fiber-pigtailed OPA module, we observed over 6.3-dB CW squeezed light from the DC component to 6-THz sideband frequencies even in a fully fiber-closed optical system. From the squeezing-level measurement, we estimate the overall optical loss of the OPA system as 21%, which is close to the loss of the OPA system without the pump light. This means that the waveguide loss is stable even with the pump light of over hundreds of milliwatts. By excluding the losses due to modularization and detection, the squeezing level at the output of the PPLN waveguide is estimated as 10.3 dB.

Equation (1) describes the deterioration of squeezing levels due to optical loss $\rho$.[12]

$$R_{\pm} = \rho + (1-\rho)exp(\pm 2\sqrt{aP}), \qquad (1)$$

where $R_-$ and $R_+$ are squeezing and anti-squeezing levels, respectively, $a$ is second-harmonic (SH) conversion efficiency of the OPA, and $P$ is pump power in the OPA. From Eq. (1), it is easily understood that a higher squeezing level requires not only low optical loss but also higher pump power injection. The strong pump light leads to high parametric gain, but it often causes various problems in an $\chi^{(2)}$ waveguide OPA. For example, the electrical field of strong light causes spatial distribution of electrons in the $\chi^{(2)}$ media, which results in spatial fluctuation of the refractive index according to an electro-optic effect. This phenomena, known as the photorefractive effect, causes scattering of propagating light.[26] Strong light often also causes photo-induced light absorption, such as green- (or visible-) induced infrared absorption.[28,29] Moreover, for achieving higher squeezing level, it is important to suppress excitation of higher order modes because anti-squeezed noise of higher-order modes reduces the squeezing level.[17]

Here, we fabricated a low-loss quasi-single-mode ZnO-doped PPLN waveguide by direct bonding[31] and mechanical sculpturing methods[38] to achieve a high squeezing level. Unlike a dry-etched waveguide, which usually has large sidewall roughness due to redeposition of byproducts during the etching process, a sculptured waveguide has smooth sidewalls. Furthermore, the high-power-pump durability of the directly bonded ZnO-doped waveguide provides higher parametric amplification gain without any pump-induced deterioration.[17,30,31] In addition, we designed the waveguide structure to allow the existence of only two modes for squeezed light. Since even-number-order modes for squeezed light have an asymmetric shape, they hardly interact with the Gaussian-like-shaped first-order mode of the pump. This means that the amplitude of the generated second-ordered mode is negligibly small as long as we inject Gaussian shaped pump light. Therefore, we can neglect contamination of the anti-squeezed light from higher order modes by using a waveguide with a secondary or lower mode. Figure 1(a) shows a schematic of the fabrication of a mechanically sculptured waveguide. First, we directly bonded a ZnO-doped PPLN layer onto a LiTaO$_3$ substrate. The thickness of the PPLN layer is about 8.0 μm. The waveguide was sculptured with a mechanical saw, which is made of a diamond polisher and resin. Figure 1(b) shows a cross-sectional view of a fabricated waveguide, whose width is about 8.6 μm. A sidewall angle of this waveguide is about 87°, which is higher than that of typical dry etched waveguide.[17] This helps more efficient optical coupling to symmetrical modes from optical fibers. The waveguide length is 45 mm. The periodically poled pattern was fabricated by an electrical poling process[31] with the poling period of about 18 μm. Both the input and output end-faces of the waveguide were covered with an anti-reflection (AR) coating for both the 1545- and 773-nm wavelengths of light. Figure 1(c) shows measured waveguide loss. The 45-mm-long waveguide shows about 7% transmission loss for the 1545-nm wavelength of light. Namely, the waveguide has transmission loss of 0.1 dB/cm at most. This value is lower than 0.3 dB/cm reported in Ref. 33-35, which was obtained for typical dry-etched waveguides. The second-harmonic generation (SHG) coefficient at the phase-matched wavelength is 820 %/W, as shown in Fig. 1(d), which was acquired under 3-mW light injection with a wavelength of 1545.3 nm. Therefore, the efficiency per unit length is 40 %/W·cm$^2$, which is lower than the reported value of 370 %/W·cm$^2$ in Ref. 18 mostly due to its large core size. The slight difference of the SHG curve from an ideal sinc-squared shape would be caused by a structural perturbation and a birefringence inhomogeneity of the waveguide. To achieve higher SHG conversion efficiency, we need to fabricate more uniform and low-loss waveguide with smaller core.

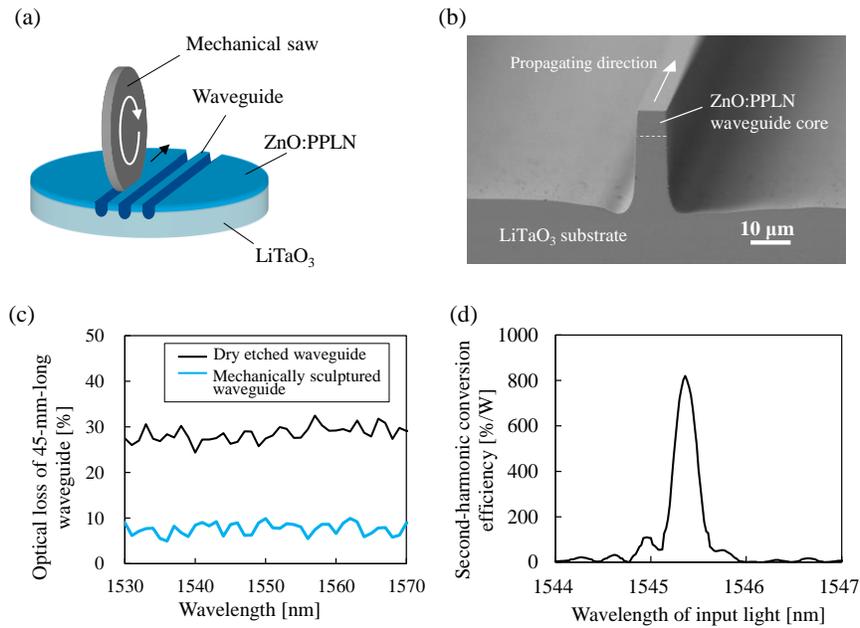

FIG. 1. (a) Schematic of how we fabricated a mechanically sculptured PPLN waveguide. (b) Cross-sectional view of the PPLN waveguide taken by scanning electron-beam microscopy. (c) Measured waveguide loss spectra for a dry etched waveguide and mechanically sculptured waveguide with length of 45 mm. (d) Second-harmonic conversion efficiency of fabricated PPLN waveguide.

For practical use, a modularized squeezer is preferable as a quantum light source[24,25] because, like a laser used as a coherent light source, it contributes reducing the maintenance and increasing the portability of optical systems for out-of-lab applications. Here, we modularized the fabricated PPLN waveguide into a fiber-pigtailed squeezer as shown in Fig. 2(a). This module shows overall optical loss of about 1.0 dB, namely 21%, for the communications wavelength band. The module houses the mechanically sculptured PPLN waveguide, six lenses, four dichroic mirrors, and four pigtailed optical fibers, as shown in Fig. 2(b). The waveguide is put on a copper plate. To maintain the phase-matching condition at any environmental temperature, we put a Peltier device under the copper plate to keep the temperature of overall the waveguide. The dichroic mirrors combine the second-harmonic-pump and telecommunication-band light on one optical axis or separate them. To minimize optical coupling loss between the waveguide and pigtailed optical fibers, we use AR-coated aspherical lenses. Mode-mismatch loss between the optical fiber for communications band light and the PPLN waveguide was designed to be below 1%. The loss of the optical components, such as the mirrors and lenses, was measured as 6% in total for one input (or output) side. Therefore, with including the measured waveguide loss of 7%, the overall optical loss of the module is expected to be about 18% ideally. The slight difference from the measured loss of 21% is thought be caused by imperfect alignment inside the module and mode-mismatch due to waveguide fabrication error. By excluding 7% of the waveguide loss from 21% of the measured total insertion loss of the OPA system, the loss caused by the modularization is estimated as 8% for each input (or output) side.

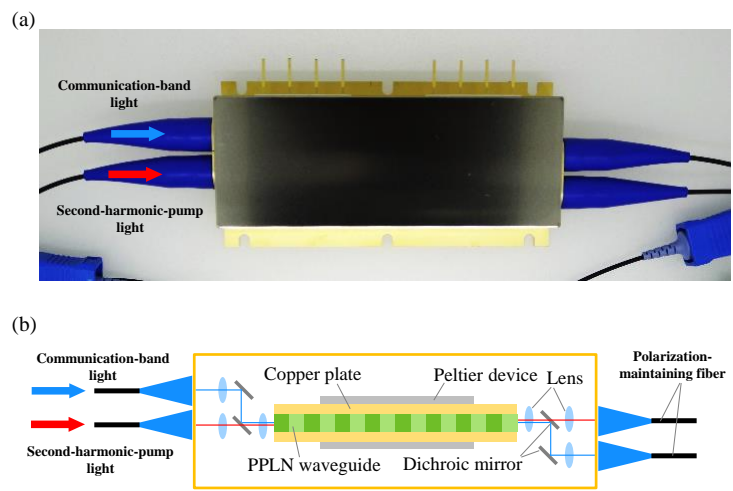

FIG. 2. (a) Overview of fabricated OPA module. (b) Schematic of the inside of the fabricated OPA module. The PPLN waveguide length is 45 mm.

To measure THz-order broadband squeezed light, we used an all-optical quadrature measurement technique with the use of another OPA[42,43] instead of conventional balanced homodyne measurement.[44] This is because the bandwidth of conventional balanced homodyne measurement is limited to at most gigahertz-order range due to its electrical circuits.[45] On the other hand, the bandwidth of the all-optical detection technique is in the THz-order range because the OPA amplifies one optical quadrature while attenuating the other quadrature in the optical region. The amplitudes of the amplified light contain information on the original quadrature amplitudes. With the use of a similar technique, optical phase detection with a 6-dB-suppressed shot-noise limit was demonstrated with a nonwaveguide OPA pumped by pulsed light in 2021.[46] In the case of squeezing-level measurement, we only need to measure optical intensities of amplified light.[43] According to Ref. 43, measured squeezing levels $R'_\pm$ are described by using detected optical intensities as follows:

$$R'_\pm = \frac{I_\pm}{I_0} = \frac{1}{1+G^2} R_\mp + \frac{G^2}{1+G^2} R_\pm. \qquad (2)$$

Here, $I_0$ is the intensity of the amplified vacuum noise, $I_+$ is the intensity of the amplified anti-squeezed noise, $I_-$ is the intensity of the amplified squeezed noise, and $G$ is the gain of the OPA. In this method, the deterioration of the squeezing level occurs only in the optical region in front of the amplifying OPA because the amplified optical signal is tolerant to optical loss.[46] Therefore, to observe high-level squeezing, we need a low-loss OPA in terms of not only squeezed-light generation but also its detection.

Figure 3 shows the experimental setup for measuring squeezing levels from optical intensities of light amplified by the fabricated OPA. This setup is almost same as the one in Ref. 43. CW laser light with wavelength of 1545.3 nm (194.0 THz) is emitted from a single-frequency fiber laser (NKT Photonics, BASIK module). The OPAs are used for squeezed-light generation (OPA1) and for amplification of the generated squeezed light (OPA2). The power of the injected pump for each OPA is controlled by the injection current for an EDFA (Keopsys, CEFA-C-PB-HP) placed before a PPLN-waveguide frequency doubler (NTT Electronics, WH-0772-000-F-B-C). The amplified squeezed light was divided into two paths. One is for measuring its intensity with an optical spectrum analyzer (OSA) (YOKOGAWA, AQ6340). The other is for relative phase locking between the squeezed light and second-harmonic pump light in OPA2. The gain of OPA2 was measured as 20 dB ($G = 100$). Therefore, according to Eq. (2), we can reduce the effect of the original anti-squeezing level to about 1/10000 in the squeezing-level detection.

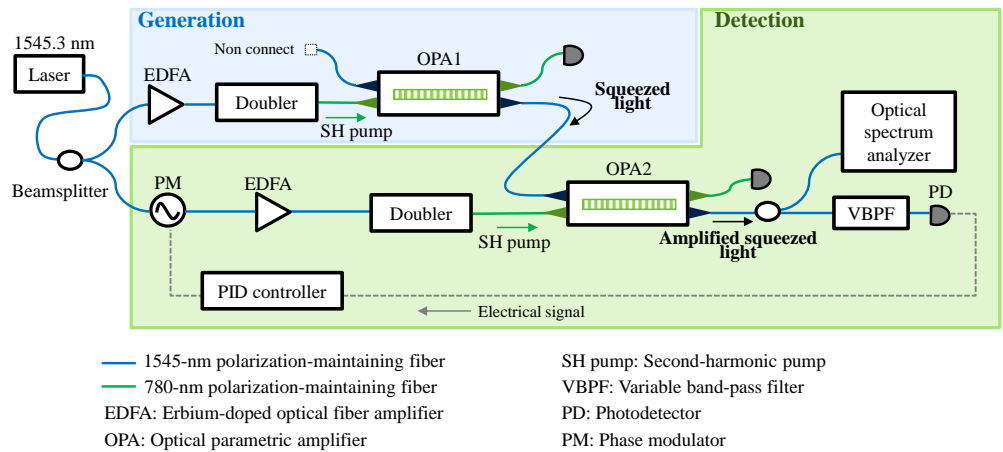

FIG. 3. Schematic diagram of the experimental setup for detection of squeezed light from a PPLN waveguide module (OPA1). The generated squeezed light is amplified in another PPLN waveguide module (OPA2).

Figure 4(a) shows measured optical intensities as a function of sweep time at the wavelength of 1545.0 nm, which corresponds to a 40-GHz sideband frequency. The gray and red curves are vacuum and squeezed noise levels without phase locking. This figure also shows squeezed and anti-squeezed noise levels with phase locking. The vacuum noise is detected without pump injection for OPA1. The phase-locked squeezed noise level is 6.4 dB below the vacuum noise level.

Figure 4(b) shows observed squeezing and anti-squeezing levels as a function of injected SH-pump light intensity for OPA1. The theoretical curve is described by Eq. (1) with fitting parameters of SHG coefficient $a$ and effective optical loss $\rho$. The obtained SHG coefficient is 823 %/W, which corresponds to the value of 820 %/W as the measured value shown in Fig. 1(d). This indicates that our OPA works without any deterioration, such as that due to the photorefractive effect, under pump injection of hundreds of milliwatts. On the other hand, the estimated optical loss is about 21%. The loss includes the waveguide loss and coupling loss between optical fibers and waveguides. Assuming that the squeezed light is generated in the middle of OPA1 and detected in the middle of OPA2,[43] the total loss becomes the same value as the loss of one module of 1.0 dB, namely 21%. This coincidence is also evidence that there is little influence of pump-induced deterioration. Here, with considering 8% loss for the output side of OPA1 and 8% for the input side of OPA2, total detection loss is calculated as 15% ($\cong$ 1-0.92·0.92). By subtracting this loss, the squeezing level soon after the PPLN waveguide is estimated as 10.3 dB at 1545.0 nm. Therefore, by using more efficient detection technique, we will observe higher squeezing level than 6.4 dB.

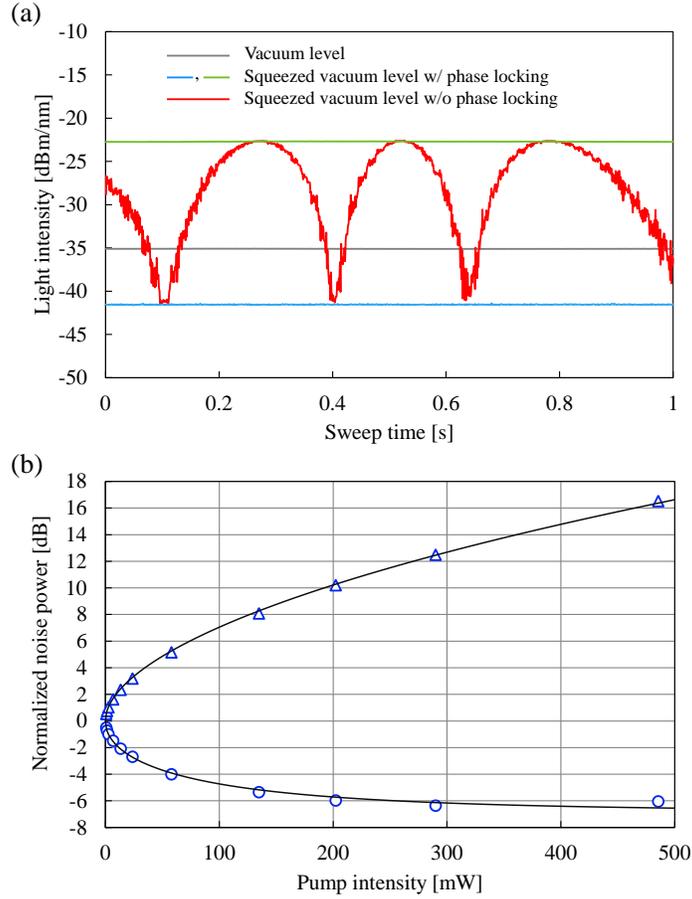

FIG. 4. (a) Vacuum and squeezed vacuum noise levels detected by an optical spectrum analyzer in zero-span mode at the wavelength of 1545.0 nm, which corresponds to a 40-GHz sideband. The resolution bandwidth was set at 0.2 nm. (b) Pump power dependence of squeezed and anti-squeezed noise levels.

To evaluate the bandwidth of the squeezed light, we measured optical spectra of the amplified vacuum and phase-locked squeezed vacuum states of light, as shown in Fig. 5(a). The blue and orange curves are squeezed and anti-squeezed at the center frequency, respectively. The gray curve is the vacuum level, which was detected without pump injection for OPA1. The optical quadrature squeezing is observed in the wavelength range from 1490 to 1600 nm, which means that our waveguide OPA generates squeezed light with THz-order bandwidth. There are ripples in the spectra due to chromatic dispersion in the pigtailed optical fibers of the OPAs. By using dispersion-compensating fiber, we can broaden the phase-maintained bandwidth of the squeezed light.[43]

To measure the squeezing level at various sideband frequencies, we observed the light intensities with zero-span measurement at the wavelengths of 1587.0 nm and 1595.0 nm, as shown in Fig. 5(b) and Fig. 5(c), respectively. Even at 1595.0 nm (corresponding to a 6.0 THz sideband), the squeezing level was measured as 6.3 dB. The vacuum noise level at 1595 nm is lower than that at 1545.3 nm as the center wavelength due to the gain reduction of OPA2 with departure from the center frequency. However, the measured squeezing level is maintained from the DC component to the 6.0-THz sideband frequency because the gain of OPA2 was sufficient for squeezing-level measurement in this experiment. These results mean that our waveguide squeezer generates 6.3-dB 6-THz-bandwidth squeezed light even in a fully fiber-closed optical system.

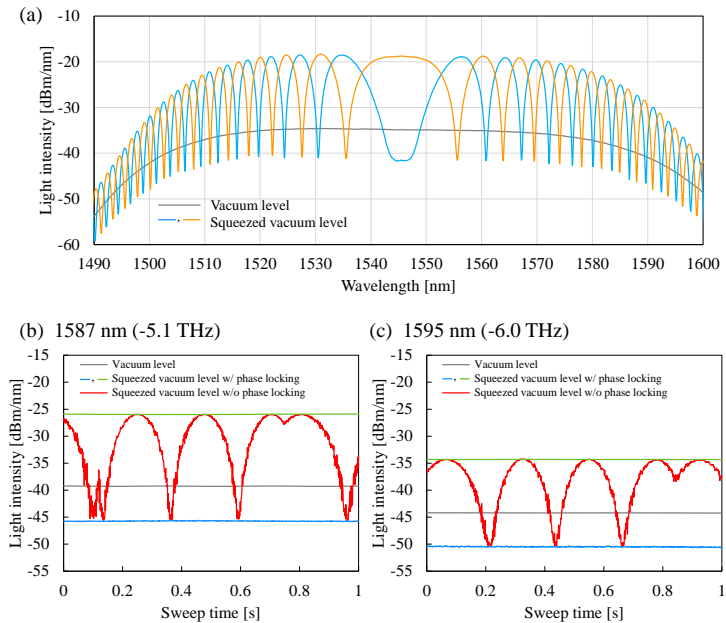

FIG. 5. (a) Result of all-optical phase-sensitive detection obtained with an optical spectrum analyzer. The resolution bandwidth was set at 0.2 nm. (b), (c) Vacuum and squeezed noise levels detected by an optical spectrum analyzer in zero-span mode at the wavelengths of (b) 1587.0 nm and (c) 1595.0 nm.

In summary, to generate CW broadband high-level squeezed light, we fabricated a low-loss quasi-single-mode ZnO-doped PPLN waveguide using mechanical polishing process. Thanks to its smooth sidewalls, the waveguide shows 7% optical propagation loss at a 1545-nm wavelength of light with a waveguide length of 45 mm. The ZnO-doped PPLN waveguide core was directly bonded onto a lithium tantalate substrate for high durability against high-power-pump injection to suppress various pump-induced phenomena. Furthermore, the quasi-single-mode waveguide helps to suppress contamination of the anti-squeezed component of higher order modes. By using the fabricated waveguide assembled as a fiber-pigtailed OPA module, we observed over 6.3-dB CW squeezed light from the DC component to 6-THz sideband frequencies even in a fully fiber-closed optical system. From the squeezing-level-measurement, we estimate the overall optical loss of the OPA system as 21%, which is close to the value of the measured loss of the OPA system without the pump light. This means that the waveguide loss is stable even with pump power of hundreds of milliwatts. By excluding 15% loss caused by the modularization and detection, the squeezing level at the output of the PPLN waveguide is estimated as 10.3 dB. This represents great progress toward the realization of high-speed large-scale fault-tolerant MBQC. Furthermore, the fiber-pigtailed configuration is useful for practical optical quantum computing thanks to its compatibility with various reliable optical components. It promises maintenance-free and portable optical systems for out-of-lab applications.

## ACKNOWLEDGEMENT

This work was partially supported by JST-Moonshot Research and Development Program (JPMJMS2064), KAKENHI (18H05297) of Japan Society for the Promotion of Science (JSPS), ALPS and FoPM of Ministry of Education, Culture, Sports, Science and Technology (MEXT), and The University of Tokyo Foundation. The author acknowledges Kei Watanabe, Takushi Kazama, Koji Enbutsu, Osamu Tandanaga, and Yoshiki Nishida for their useful comments on the manuscript.

## DATA AVAILABILITY

The data that support the findings of this study are available from the corresponding author upon reasonable request.

## REFERENCES

1. S. Takeda and A. Furusawa, "Toward large-scale fault-tolerant universal photonic quantum computing," APL Photonics **4**, 60902 (2019).

2. J. E. Bourassa, R. N. Alexander, M. Vasmer, A. Patil, I. Tzitrin, T. Matsuura, D. Su, B. Q. Baragiola, S. Guha, G. Dauphinais, K. K. Sabapathy, N. C. Menicucci, and I. Dhand, "Blueprint for a Scalable Photonic Fault-tolerant quantum computer," Quantum **5**, 392 (2021).

3. N. C. Menicucci, P. van Loock, M. Gu, C. Weedbrook, T. C. Ralph, and M. A. Nielsen, "Universal quantum computation with continuous-variable cluster states," Phys. Rev. Lett. **97**, 110501 (2006).

4. S. Yokoyama, R. Ukai, S. C. Armstrong, C. Sornphiphatphong, T. Kaji, S. Suzuki, J. I. Yoshikawa, H. Yonezawa, N. C. Menicucci, and A. Furusawa, "Ultra-large-scale continuous-variable cluster states multiplexed in the time domain," Nat. Photonics **7**, 982 (2013).

5. W. Asavanant, Y. Shiozawa, S. Yokoyama, B. Charoensombutamon, H. Emura, R. N. Alexander, S. Takeda, J.-I. Yoshikawa, N. C. Menicucci, H. Yonezawa, and A. Furusawa, "Generation of time-domain-multiplexed two-dimensional cluster state," Science **366**, 373 (2019).

6. A. I. Lvovsky, P. Grangier, A. Ourjoumtsev, V. Parigi, M. Sasaki, and R. T. Tualle-Brouri, "Production and application of non-Gaussian quantum states of light," arXiv:2006.16985.

7. W. Asavanant, B. Charoensombutamon, S. Yokoyama, T. Ebihara, T. Nakamura, R. N. Alexander, M. Endo, J. Yoshikawa, N. C. Menicucci, H. Yonezawa, and A. Furusawa, "One-hundred step measurement-based quantum computation multiplexed in the time domain with 25 MHz clock frequency," arXiv:2006.11537.

8. K Fukui, W Asavanant, and A Furusawa, "Temporal-mode continuous-variable three-dimensional cluster state for topologically protected measurement-based quantum computation," Phys. Rev. A **102**, 032614 (2020).

9. K. Wakui, Y. Eto, H. Benichi, S. Izumi, T. Yanagida, K. Ema, T. Numata, D. Fukuda, M. Takeoka, and M. Sasaki, "Ultrabroadband direct detection of nonclassical photon statistics at telecom wavelength," Sci. Rep. **4**, 4535 (2014).

10. K. Yoshino, T. Aoki, and A. Furusawa, "Generation of continuous-wave broadband entangled beams using periodically poled lithium niobate waveguides," Appl. Phys. Lett **90**, 041111 (2007).

11. M. Pysher, R. Bloomer, C. M. Kaleva, T. D. Roberts, P. Battle, and O. Pfister, "Broadband amplitude squeezing in a periodically poled $KTiOPO_4$ waveguide," Opt. Lett. **34**, 256 (2009).

12. D. K. Serkland, M. M. Fejer, R. L. Byer, and Y. Yamamoto, "Squeezing in a quasi-phase-matched $LiNbO_3$ waveguide," Opt. Lett. **20**, 1649 (1995).

13. L. Ldezma, R. Sekine, Q. Guo, R. Nehra, S. Jahani, and A. Marandi, "Intense optical parametric amplification in dispersion engineered nanophotonics lithium niobate waveguides," arXiv:2104.08262.

14. J. Riemensberger, J. Liu, N. Kuznetsov, J. He, R. N. Wang, and T. Kippenberg, "Photonic chip-based continuous-travelling-wave parametric amplifier," arXiv:2107.09582.

15. T. Kishimoto, K. Inafune, Y. Ogawa, N. Sekine, H. Murai, and H. Sasaki, "Periodically poled $LiNbO_3$ ridge waveguide with 21.9 dB phase-sensitive gain by optical parametric amplification," in OFC 2017 OSA Technical Digest (Optical Society of America, 2017), paper W2A.12.

16. T. Kashiwazaki, K. Enbutsu, T. Kazama, O. Tadanaga, T. Umeki, and R. Kasahara, "Over-30-dB phase-sensitive amplification using a fiber-pigtailed PPLN waveguide module," in Nonlinear Optics 2019 OSA Technical Digest (Optical Society of America, 2019), paper NW3A.2.

17. T. Kashiwazaki, N. Takanashi, T. Yamashima, T. Kazama, K. Enbutsu, R. Kasahara, T. Umeki, and A. Furusawa, "Continuous-wave 6-dB-squeezed light with 2.5-THz-bandwidth from single-mode PPLN waveguide," APL Photonics **5**, 036104 (2020).

18. S. Kurimura, Y. Kato, M. Maruyama, Y. Usui, and H. Nakajima, "Quasi-phase-mached adhered ridge waveguide in $LiNbO_3$," Appl. Phys. Lett. **89**, 191123 (2006).

19. R. Kou, S. Kurimura, K. Kikuchi, A. Terasaki, H. Nakajima, K. Kondou, and J. Ichikawa, "High-gain, wide-dynamic-range parametric interaction in Mg-doped $LiNbO_3$ quasi-phase-matched adhered ridge waveguide," Opt. Exp. **19**, 11867 (2011).

20. K. Fukui, A. Tomita, A. Okamoto, and K. Fujii, "High-threshold fault-tolerant quantum computation with analog quantum error correction," Phys. Rev. X **8**, 021054 (2018).

21. K. Fukui "High-threshold fault-tolerant quantum computation with the GKP qubit and realistically noisy devices," arXiv:1906.09767.

22. K. Noh, and C. Chamberland, "Fault-tolerant bosonic quantum error correction with the surface-Gottesman-Kitaev-Preskill code," Phys. Rev. A **101**, 012316 (2020).

23. M. V. Larsen, C. Chamberland, K. Noh, J. S. Neergaard-Nielsen, and U. L. Andersen, "A fault-tolerant continuous-variable measurement-based quantum computation architecture," arXiv:2101.03014.

24. F. Kaiser, B. Fedrici, A. Zavatta, V. D'Auria, and S. Tanzilli, "A fully guided-wave squeezing experiment for fiber quantum networks," Optica **3**, 362 (2016).

25. N. Takanashi, T. Kashiwazaki, T. Kazama, K. Enbutsu, R. Kasahara, T. Umeki, and A. Furusawa, "4-dB quadrature squeezing with fiber-coupled PPLN ridge waveguide module," J. Quantum Electron. **56**, 6000100 (2020).

26. A. M. Glass, "The photorefractive effect," Opt. Eng. **17**, 470 (1978).

27. J. Jackel, A. M. Glass, G. E. Peterson, C. E. Rice, D. H. Olson, and J. J. Veselka, "Damage-resistant $LiNbO_3$ waveguides," J. Appl. Phys. **55**, 269 (1984).

28. Y. Furukawa, K. Kitamura, A. Alexandrovski, R. K. Route, M. M. Fejer, and G. Foulon, "Green-induced infrared absorption in MgO doped $LiNbO_3$," Appl. Phys. Lett. **78**, 1970 (2001).

29. M. Imlau, H. Badorreck, and C. Merschjann, "Optical nonlinearities of small polarons in lithium niobate," Appl. Phys. Rev. **2**, 040606 (2015).

30. M. Asobe, O. Tadanaga, T. Yanagawa, H. Itoh, and H. Suzuki, "Reducing photorefractive effect in periodically poled ZnO- and MgO-doped $LiNbO_3$ wavelength converters," Appl. Phys. Lett. **78**, 3163 (2001).

31. T. Umeki, O. Tadanaga, and M. Asobe, "Highly efficient wavelength converter using direct-bonded PPZnLN ridge waveguide," IEEE J. Quantum Electron. **46**, 1206 (2010).

32. J. Singh and R. M. De La Rue, "An experimental study of in-plane light scattering in titanium diffused Y-cut $LiNbO_3$ optical waveguides," J. Light. Technol. **3**, 67 (1985).

33. L. Cai, A. Mahmoud, and G. Piazza, "Low-loss waveguide on Y-cut thin film lithium niobate: towards acousto-optic applications," Opt. Exp. **27**, 9794 (2019).

34. K. Luke, P. Kharel, C. Reimer, L. He, M. Loncar, and M. Zhang, "Wafer-scale low-loss lithium niobate photonic integrated circuits," Opt. Exp. **28**, 24452 (2020).

35. S. Y. Siew, E. J. H. Cheung, H. Liang, A. Bettiol, N. Toyoda, B. Alsehri, E. Dogheche, and A. J. Danner, "Ultra-low loss ridge waveguides on lithium niobate via argon ion milling and gas clustered ion beam smoothening," Opt. Exp. **26**, 4421 (2018).

36. G. Ulliac, V. Calero, A. Ndao, F. I. Baida, and M. P. Bernal, "Argon plasma inductively coupled plasma reactive ion etching study for smooth sidewall thin film lithium niobate waveguide application," Opt. Mater. **53**, 1 (2016).

37. H. Nagata, N. Mitsugi, K. Shima, M. Tamai, and E. M. Haga, "Growth of crystalline LiF on $CF_4$ plasma etched $LiNbO_3$ substrates," Journal of Crystal Growth **187**, 573 (1998).

38. Y. Nishida, H. Miyazawa, M. Asobe, O. Tadanaga, and H. Suzuki, "Direct-bonded QPM-LN ridge waveguide with high damage resistance at room temperature," Electron. Lett. **39**, 609 (2003).

39. M. Chauvet, F. Henrot, F. Bassignot, F. Devaux, L. Gauthier-Manuel, V. Pêcheur, H. Maillotte, and B. Dahman, "High efficiency frequency doubling in fully diced $LiNbO_3$ ridge waveguides on silicon," J. Opt. **18**, 085503 (2016).

40. N. Courjal, B. Guichardaz, G. Ulliac, J.-Y. Rauch, B Sadani, H.-H. Lu, and M.-P. Berna, "High aspect ratio lithium niobate ridge waveguides fabricated by optical grade dicing," J. Phys. D: Appl. Phys. **44**, 305101 (2011).

41. M. F. Volk, S. Suntsov, C. E. Rüter, and D. Kip, "Low loss ridge waveguides in lithium niobate thin films by optical grade diamond blade dicing," Opt. Exp. **24**, 1386 (2016).

42. Y. Shaked, Y. Michael, R. Z. Vered, L. Bello, M. Rosenbluh, and A. Pe'er, "Lifting the bandwidth limit of optical homodyne measurement with broadband parametric amplification," Nat. Commun. **9**, 609 (2018).

43. N. Takanashi, A. Inoue, T. Kashiwazaki, T. Kazama, K. Enbutsu, R. Kasahara, T. Umeki, and A. Furusawa, "All-optical phase-sensitive detection for ultra-fast quantum computation," Opt. Exp. **28**, 34916 (2020).

44. H. P. Yuen, and V. M. S. Chan, "Noise in homodyne and heterodyne detection," Opt. Lett. **8**, 177 (1983).

45. J. F. Tasker, J. Frazer, G. Ferranti, E. J. Allen, L. F. Brunel, S. Tanzilli, V. D'Auria, J. C. F. Matthews, "Silicon photonics interfaced with integrated electronics for 9 GHz measurement of squeezed light," Nat. Photon. **15**, 11 (2021).

46. G. Frascella, S. Agne, F. Khalili, and M. V. Chekhova, "Overcoming detection loss and noise in squeezing-based optical sensing," arXiv:2005.08843.